\documentclass[preprint,aps,prl,superscriptaddress,nofootinbib]{revtex4-1}
\usepackage[utf8]{inputenc}
\DeclareUnicodeCharacter{FB01}{fi}

\usepackage{amsmath}    
\usepackage{graphicx}   
\usepackage{verbatim}   
\usepackage{color}      
\usepackage{epstopdf}  
\usepackage{upgreek} 	
\usepackage[hidelinks]{hyperref}   
\usepackage{textcomp} 
\usepackage{siunitx}
\usepackage{mathtools}

\newcommand{\mumax}{MuMax$^{3}$ }
\newcommand{\figref}[1]{Fig.~\ref{#1}}

\begin{document}

\title{Chiral Spin Spirals at the Surface of \\ the van der Waals Ferromagnet Fe$_3$GeTe$_2$}
\author{Mariëlle J. Meijer}
\email{m.j.meijer@tue.nl}
\affiliation{Department of Applied Physics, Eindhoven University of Technology, P.O. Box 513, 5600 MB Eindhoven, the Netherlands}
\author{Juriaan Lucassen}
\affiliation{Department of Applied Physics, Eindhoven University of Technology, P.O. Box 513, 5600 MB Eindhoven, the Netherlands}
\author{Rembert A. Duine}
\affiliation{Department of Applied Physics, Eindhoven University of Technology, P.O. Box 513, 5600 MB Eindhoven, the Netherlands}
\affiliation{Institute for Theoretical Physics, Utrecht University, Leuvenlaan 4, 3584 CE Utrecht, the Netherlands}
\author{Henk J.M. Swagten}
\affiliation{Department of Applied Physics, Eindhoven University of Technology, P.O. Box 513, 5600 MB Eindhoven, the Netherlands}
\author{Bert Koopmans}
\affiliation{Department of Applied Physics, Eindhoven University of Technology, P.O. Box 513, 5600 MB Eindhoven, the Netherlands}
\author{Reinoud Lavrijsen}
\affiliation{Department of Applied Physics, Eindhoven University of Technology, P.O. Box 513, 5600 MB Eindhoven, the Netherlands}
\author{Marcos H. D. Guimar\~aes}
\email{m.h.guimaraes@rug.nl}
\affiliation{Department of Applied Physics, Eindhoven University of Technology, P.O. Box 513, 5600 MB Eindhoven, the Netherlands}
\affiliation{Zernike Institute for Advanced Materials, University of Groningen, Nijenborgh 4, 9747 AG Groningen, the Netherlands}

\date{\today}
\begin{abstract}
Topologically protected magnetic structures provide a robust platform for low power consumption devices for computation and data storage. Examples of these structures are skyrmions, chiral domain walls, and spin spirals. Here we use scanning electron microscopy with polarization analysis to unveil the presence of chiral counterclockwise N\'eel spin spirals at the surface of a bulk van der Waals ferromagnet Fe$_3$GeTe$_2$ (FGT), at zero magnetic field. These N\'eel spin spirals survive up to FGT's Curie temperature $T_\mathrm{C}= 220$~\si{K}, with little change in the periodicity $p=300$~\si{nm} of the spin spiral throughout the studied temperature range. The formation of a spin spiral showing counterclockwise rotation strongly suggests the presence of a positive Dzyaloshinskii-Moriya interaction in FGT, which provides the first steps towards the understanding of the magnetic structure of FGT. Our results additionally pave the way for chiral magnetism in van der Waals materials and their heterostructures.

\end{abstract} 
\maketitle
 
Magnetism in layered systems has proven to be a fertile ground for emergent magnetic phenomena.
The absence of inversion symmetry combined with large spin-orbit coupling in some of these structures can give rise to an asymmetric exchange interaction known as the Dzyaloshinskii-Moriya interaction (DMI) \cite{Bode2007ChiralAsymmetry, PhysRevLett.87.037203, Fert2017, Heinze2011SpontaneousDimensions}.
Systems with a large DMI offer a huge playground for the exploration of topologically protected magnetic structures such as skyrmions and chiral domain walls, which have dimensions in the order of tens of nm and are promising elements for low-power consumption electronics \cite{Fert2017, doi:10.1063/1.5048972}.
Long range non-collinear magnetic structures can also arise in materials with large DMI, where the magnetization continuously varies in the material in a sinusoidal fashion \cite{Bode2007ChiralAsymmetry, Fert2017, PhysRevB.88.184422, PhysRevLett.101.027201}.  
These structures, named spin spirals, carry important information on the magnetic properties of the system through their periodicity and handedness.
Moreover, spin spirals have been shown to evolve into skyrmions in the presence of a sufficiently large magnetic field for various material systems \cite{Schmidt_2016, Fert2017, Herve2018}. 

The recent discovery of magnetic ordering in van der Waals (vdW) materials down to the monolayer limit \cite{Gong2017, Huang2017Layer-dependentLimit} has opened a new direction in the field of two-dimensional materials, allowing researchers to explore magnetism in lower dimensions in simple crystal systems \cite{Burch2018MagnetismMaterials, Gong2019Two-dimensionalDevices, Gibertini2019MagneticHeterostructures, Mak2019ProbingMaterials}.
Particularly, the metallic vdW ferromagnet Fe$_3$GeTe$_2$ (FGT) shows large out-of-plane magnetic anisotropy and high Curie temperature ($T_\mathrm{C}$ = 220 K) \cite{Leon-Brito2016MagneticFe3GeTe2, Fei2018Two-dimensionalFe3GeTe2}, which can be pushed above room temperature upon doping \cite{Deng2018Gate-tunableFe3GeTe2} or patterning \cite{Li2018Patterning-InducedTemperature}.
The large out-of-plane magnetic anisotropy of FGT indicates a high spin-orbit coupling and opens the possibility to form interesting magnetic textures, such as skyrmions or spin spirals.
These magnetic textures in vdW magnets are still largely unexplored and are currently a topic to which significant research efforts are devoted \cite{Wu2019Neel-typeHeterostructure,Wang2019DirectNanolayers,park2019observation,Zhong2020}.
Unveiling these spin structures in two-dimensions can give a significant push towards a deeper understanding of magnetism in lower dimensions, along with the prospect of using vdW magnets for future applications.

\begin{figure*}
\centering
\includegraphics{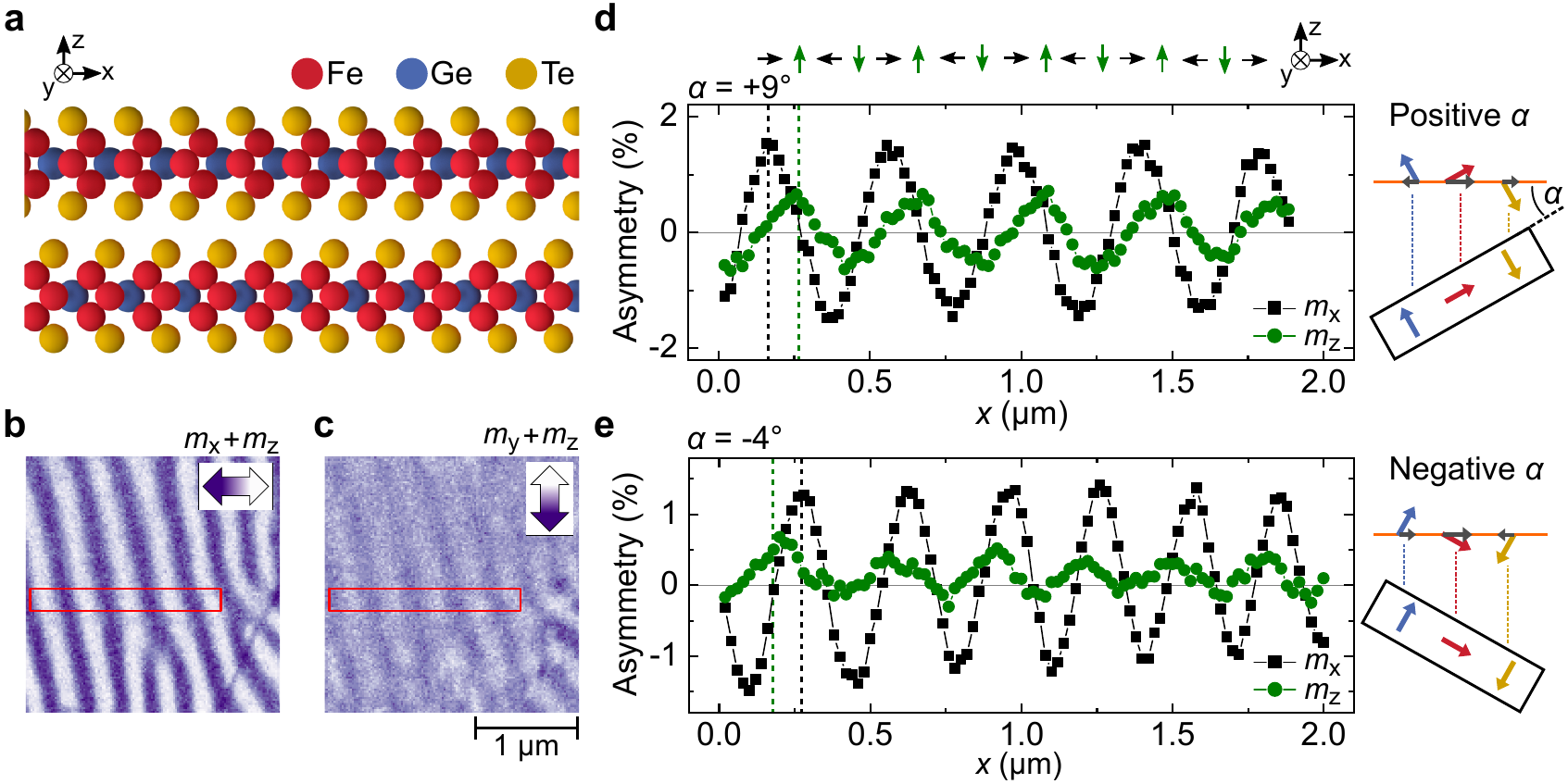}
\caption{\label{fig:figure1} Spin spirals at the surface of a $d=185$~\si{nm} thick FGT flake at $T=150$~\si{K}. \textbf{a)} Crystal structure of two FGT layers.
\textbf{b), \textbf{c)}} SEMPA images measured at the surface of FGT for $\alpha=+9^\circ$. Panel \textbf{b)} shows $m_\mathrm{x}$ contrast and panel \textbf{c)} $m_\mathrm{y}$ for the exact same area, with the color scale (in arbitrary units) indicated by the arrows in the top right-hand corner.
Additionally, in both SEMPA images an out-of-plane magnetization $m_\mathrm{z}$ can be present, which is adjustable in panel \textbf{c)} only.
\textbf{d), e)} Averaged magnetization profiles obtained from SEMPA measurements for the same area. In black and green we depict the average magnetization profile in the red rectangle of panels \textbf{b)} and \textbf{c)}, respectively. 
The sample tilt, illustrated on the right, was $\alpha=+9^\circ$ in panel \textbf{d)} and $\alpha=-4^\circ$ in panel \textbf{e)}.
The phase shift reverses from $+\pi/2$ in panel \textbf{d)} to $-\pi/2$ in panel \textbf{e)}, which is expected for a $m_\mathrm{z}$ magnetization contrast in the SEMPA image in panel \textbf{c)}. Overall, we observe a counterclockwise rotating Néel spin spiral as is indicated schematically by the arrows above panel \textbf{d)}. 
}      
\end{figure*}
 
In this letter, we image the magnetic texture at the surface of FGT to identify the underlying interactions.
We reveal the presence of a spin spiral rotating out-of-plane in a counterclockwise fashion by scanning electron microscopy with polarization analysis (SEMPA) \cite{UNGURIS2001167,Oepen2005,Koike2013Spin-polarizedMicroscopy}.
The stabilization of these magnetic textures indicates that a positive DMI is present in FGT. 
Additionally, from temperature dependent measurements we find that the periodicity of the magnetization textures remains constant in the studied temperature range from $60$~\si{K} up to $T_\mathrm{C}$, which is unexpected for the large temperature-dependent anisotropy that is reported for these systems \cite{Tan2018HardFe3GeTe2}. 
These observations allow for a further understanding of the FGT magnetic structure, paving the way for chiral magnetism using vdW materials.

Our samples are obtained by mechanical exfoliation of a bulk FGT crystal on a Si wafer. 
The sample preparation was performed in high vacuum, with pressures lower than $10^{-7}$~\si{mbar} to avoid oxidation of the exfoliated FGT crystals. 
A dusting layer of Co ($0.3$~\si{nm}) was deposited using sputtering deposition and the samples were then loaded into the SEMPA microscope chamber while keeping the sample in ultra-high vacuum. 
The Co dusting layer was found to enhance the SEMPA signal while maintaining the same magnetic pattern as in pristine flakes \cite{doi:10.1063/1.347250,Lucassen2017ScanningTextures} and this is discussed in more detail in Supplementary Information Section SII.
Additional information on the sample fabrication and AFM scans of the flake can be found in the Methods section and Supplementary Information Section SI. 

A side-view of the crystal structure of FGT is schematically depicted in \figref{fig:figure1}a. 
The individual FGT layers are arranged in an AB-stacking, where each layer is rotated by 180$^{\circ}$ around the out-of-plane ($\mathrm{z}$-) axis with respect to the adjacent layers. 
FGT has a space symmetry group $\mathrm{P6_{3}/mmc}$, with the inversion symmetry point located in the space between the layers \cite{Leon-Brito2016MagneticFe3GeTe2}. 
The magnetic properties of FGT become apparent when cooling down the sample below $T_\mathrm{C}=220$~\si{K} and a perpendicular magnetic anisotropy along the $\mathrm{z}$-axis is found \cite{Leon-Brito2016MagneticFe3GeTe2, Fei2018Two-dimensionalFe3GeTe2}. 

We use SEMPA to obtain vectorial information on the surface magnetization of FGT \cite{UNGURIS2001167,Oepen2005,Koike2013Spin-polarizedMicroscopy}. 
In SEMPA, the detector for the secondary electrons in a regular scanning electron microscope is modified to provide spin sensitivity.
This is done by accelerating spin-polarized secondary electrons, emitted by the sample, towards a W(100) target. Depending on the spin polarization direction, the secondary electrons are scattered from the target to different diffraction spots. 
The difference in intensities between these diffraction spots provides a quantitative measurement of the in-plane spin polarization of the secondary electrons coming from the sample. 
The lateral spatial resolution of our system is about $30$~\si{nm} \cite{PhysRevLett.123.157201, PhysRevLett.124.207203} and due to the high surface sensitivity of SEMPA we only probe the magnetic texture of the top FGT layer (see Supplementary Information Section SI for details) \cite{Koike2013Spin-polarizedMicroscopy}.
\figref{fig:figure1}b and c show SEMPA images of the surface of a $d=185$~\si{nm} thick FGT flake (flake A) at $T=150$~\si{K}. 
Both images are measured simultaneously and probe the exact same area of the flake. \figref{fig:figure1}b shows magnetization contrast in the $\mathrm{x}$-direction ($ m_\mathrm{x}$) and \figref{fig:figure1}c $m_\mathrm{y}$ contrast, as indicated by the arrows in the top right-hand corner. 
A strong magnetization contrast is present in \figref{fig:figure1}b and a vertical stripe-like pattern is observed, revealing an alternating in-plane magnetization from left to right. 
Only a slight magnetization contrast is observed in \figref{fig:figure1}c, but a similar vertical stripe-like pattern is present.

Even though SEMPA is in principle only sensitive to the in-plane magnetization component, we are able to detect the out-of-plane direction through a projection technique \cite{Lucassen2017ScanningTextures}.
Here, we tilt the sample by an angle $\alpha$ with respect to the measurement axis, as is schematically depicted on the right side of \figref{fig:figure1}d and e. 
It results in an adjustable mixing of the out-of-plane magnetization $m_\mathrm{z}$ component in the $m_\mathrm{y}$ channel. As we will demonstrate later on, the main component in the $m_\mathrm{y}$ SEMPA image is given by the out-of-plane $m_\mathrm{z}$ contrast.
We note that a contribution from $m_\mathrm{z}$ can also be added to the $m_\mathrm{x}$ signal by an accidental tilt from the sample mount in that direction. However, we expect this contribution to be small as discussed in Supplementary Information Section SII. 

The  spatial variation of the magnetization on the surface of FGT can be better quantified by averaging the signal along the vertical direction in the region highlighted by the red rectangles in \figref{fig:figure1}b and c. 
The averaged signal for \figref{fig:figure1}b and c are shown in black and green in \figref{fig:figure1}d, respectively, where a positive tilting angle of $\alpha=+9^\circ$ was used.
Here, the magnetization contrast, or asymmetry, is plotted as a function of position where a positive asymmetry corresponds to the light purple coloring in the SEMPA images.
A  sinusoidal variation in the magnetization contrast is clearly observed in both data sets with the same periodicity but different amplitudes. Moreover, we find that the two data sets are phase shifted by $+\pi/2$, indicating a continuous spatial change in the magnetization direction.

The higher amplitude for the $m_\mathrm{x}$ signal is expected if the signal in the $m_\mathrm{y}$ detector only measures a projection of the out-of-plane magnetization. 
We confirm this by disentangling the in-plane ($m_\mathrm{y}$) and out-of-plane ($m_\mathrm{z}$) magnetic component in the SEMPA image shown in \figref{fig:figure1}c by performing sample-tilt controlled experiments. 
When we vary $\alpha$ from positive to negative values, the projection of the $m_\mathrm{z}$ signal changes sign, whereas the $m_\mathrm{x}$ and $m_\mathrm{y}$ magnetization remains (approximately) constant. This is schematically illustrated in the insets on the right side of \figref{fig:figure1}d and e. 
We expect to find the same behavior for the phase shift: upon a sign change of $\alpha$ the phase shift reverses (from $+\pi/2$ to $-\pi/2$) if the magnetization contrast is out-of-plane ($m_\mathrm{z}$), and it remains constant if the magnetization contrast is in-plane ($m_\mathrm{y}$). 
In \figref{fig:figure1}e we show the magnetization profile for the same region as in \figref{fig:figure1}d, but with a negative tilt angle of $\alpha=-4^\circ$. 
We clearly observe that the black and green data sets are now phase shifted by $-\pi/2$, which indicates that the signal in \figref{fig:figure1}c primarily consists of $m_\mathrm{z}$ contrast. 

When we combine the magnetization profiles of \figref{fig:figure1}d (and e) we are able to reconstruct the magnetic texture in the top layer of our FGT flake. 
This is depicted schematically by the arrows on top of \figref{fig:figure1}d and illustrates a magnetization that continuously rotates in the $\mathrm{xz}$-plane.
The reconstruction therefore reveals the presence of a counterclockwise rotating N\'eel spin spiral with a period of  $p=407$~\si{nm} on the surface of FGT, which is rather surprising as will be explained below. 

\begin{figure}
\centering
\includegraphics{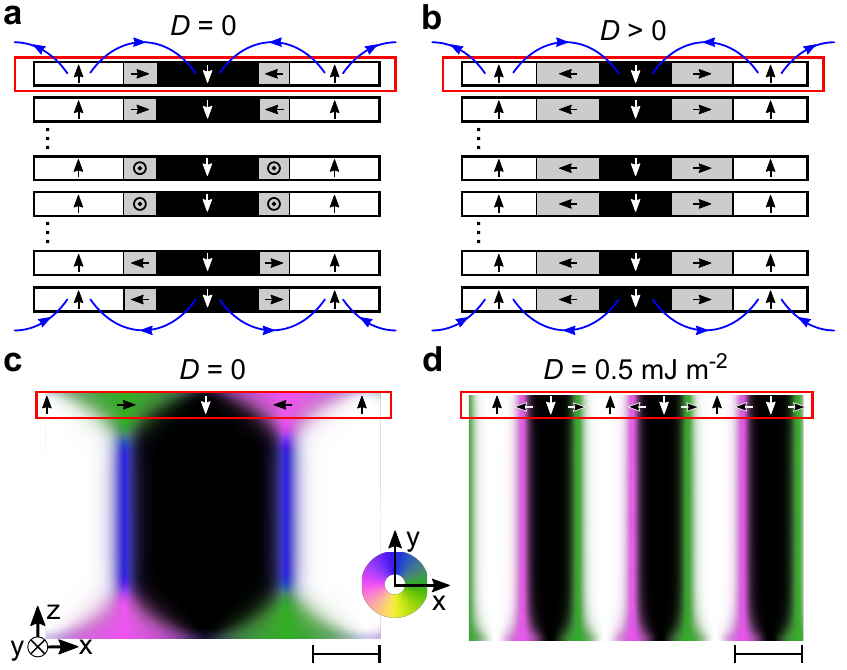}
\caption{\label{fig:figure2} \textbf{a),b)} Schematic representation of the magnetic texture in FGT, showing a side-view of six FGT layers. The white and black area correspond to up and down magnetized domains and the grey areas correspond to an in-plane magnetization, with the arrows denoting the direction. The blue arrows indicate the dipolar fields. With SEMPA only the top layer of FGT is imaged, as is indicated by the red highlighted area. In panel \textbf{a)} the DMI is zero and the magnetization in the domain walls aligns with the dipolar fields resulting in clockwise N\'eel domain wall in the top layer, Bloch walls in intermediate layers and a counterclockwise N\'eel wall in the bottom FGT layer. In panel \textbf{b)} the DMI is positive in each layer and large compared to the dipolar fields. The magnetic texture consists of an out-of-plane rotating spin spiral in a counterclockwise direction throughout all FGT layers. \textbf{c),d)} Side-view of micromagnetic simulation results for 128 layers of FGT and $K=40$~\si{kJ ~m^{-3}}. The in-plane magnetization direction is indicated by the color wheel in the $xy$-plane. In \textbf{c)} $D=0$ and in \textbf{d)} $D=0.5$~\si{mJ ~m^{-2}} and for the top FGT layer a clockwise and counterclockwise rotating spin texture is found, respectively. The scale bars on the lower right hand side indicate $25$~\si{nm}.} 
\end{figure} 

In the following we take a closer look at the interactions at play in FGT to understand in more detail why the formation of the counterclockwise spin spiral is peculiar and in addition indicates the presence of a positive DMI. 
FGT is known to exhibit a strong ferromagnetic exchange stiffness \cite{Leon-Brito2016MagneticFe3GeTe2}, and a strong (temperature-dependent)  perpendicular magnetic anisotropy \cite{Tan2018HardFe3GeTe2}. 
In \figref{fig:figure2}a we schematically depict the expected magnetic texture when including these interactions without a DMI ($D=0$) and it consists of magnetic domains separated by narrow domain walls. 
The out-of-plane magnetized domains are indicated in white and black for up and down domains, respectively, and the domains are aligned for each FGT layer due to the interlayer exchange interaction and dipolar stray fields (blue arrows).
In grey we indicate the domain walls, where the magnetization rotates in-plane.
As depicted in \figref{fig:figure2}a, the magnetization in the domain walls aligns with the direction given by the dipolar fields, which results in the formation of a clockwise rotating N\'eel wall in the top FGT layer, a Bloch wall in intermediate FGT layers and a counterclockwise N\'eel wall in the bottom FGT layer. This is analogous to the spin textures found in cobalt based magnetic multilayers without a DMI \cite{Legrandeaat0415,PhysRevLett.123.157201}. Therefore, in case dipolar fields are the dominant interaction we would expect to measure a clockwise rotating spin texture with SEMPA, since the surface sensitivity of the SEMPA only probes the top FGT layer (highlighted in red). 

The discrepancy between the measured counterclockwise spin spiral discussed in \figref{fig:figure1} and the predicted clockwise rotation of the magnetization for dipolar dominated systems (\figref{fig:figure2}a) for the top FGT layer indicates that additional interactions need to be considered.
As mentioned earlier, a known interaction to influence the chirality of the magnetization is the DMI, which we propose to be present in FGT alongside recently published studies \cite{Wu2019Neel-typeHeterostructure,Wang2019DirectNanolayers,park2019observation}.
Although similar magnetic patterns have been seen in these previous works, the specific rotation direction could not be determined, and therefore the sign of the DMI term could not be resolved.
Here we identify the DMI in FGT to be positive, as it imposes the counterclockwise rotation of the magnetic spin texture. 
This is schematically depicted in \figref{fig:figure2}b.
Additionally, the DMI lowers the domain wall energy and thereby promotes the formation of spin spirals. 

We verify the validity of the schematic images depicted in \figref{fig:figure2}a and b with micromagnetic \mumax simulations \cite{Vansteenkiste2014}. 
A side view of the simulation results are depicted in \figref{fig:figure2}c and d for $K=40$~\si{kJ ~m^{-3}} and other simulation details are specified in Supplementary Information Section IV. 
In \figref{fig:figure2}c $D=0$ and the up and down domains are separated by domain walls with a varying width across the FGT thickness. 
The in-plane magnetization direction is indicated by the color wheel depicted in the lower right hand corner. 
At the surface of FGT (highlighted by the red border) a clockwise rotating spin texture is found. 
In \figref{fig:figure2}d on the other hand, $D=0.5$~\si{mJ ~m^{-2}} in each FGT layer and a counterclockwise N\'eel spin spiral in all magnetic layers is obtained. 

A lower bound of the DMI $D_\mathrm{thres}$ can be calculated from the transformation of the domain wall textures observed in \figref{fig:figure2}. 
From \figref{fig:figure2}c we find that the majority of the domain walls consists of a Bloch wall texture (indicated in blue) and upon increasing the DMI a counterclockwise N\'eel texture is stabilized for each layer in \figref{fig:figure2}d. 
Following reference \cite{PhysRevB.95.174423}, the threshold DMI value for this system is then given by:
\begin{equation}
D_\mathrm{thres}=2\mu_0 M_\mathrm{S}^2 \left(\frac{\pi^2}{d \ln(2)}+ \pi\sqrt{\frac{K+\frac{\mu_0 M_\mathrm{S}^2}{2}}{A}}\right)^{-1},
\end{equation}
with $M_\mathrm{S}$ the saturation magnetization, $d$ the thickness of the flake, $A$ the exchange stiffness and $K$ the anisotropy, which is strongly temperature-dependent for FGT. 
Reported values for the anisotropy range from $K=1.5$~\si{MJ ~m^{-3}} for bulk FGT at 5~\si{K} \cite{Leon-Brito2016MagneticFe3GeTe2} to $K=0.23$~\si{MJ ~m^{-3}} for 10~\si{nm} FGT flakes at 120~\si{K} \cite{Tan2018HardFe3GeTe2}. 
This results in a lower bound for the DMI term of $D>0.09-0.2$~\si{mJ ~m^{-2}}, respectively, using $M_\mathrm{S}=0.38$~\si{MA ~m^{-1}} and $A=1$~\si{pJ ~m^{-1}} as reported in \cite{Leon-Brito2016MagneticFe3GeTe2}. 

So far, the presented data indicates the presence of a positive DMI in FGT, but the exact origin of this DMI remains elusive. 
The inversion symmetry of FGT, as shown in \figref{fig:figure1}a in principle suggests an absence of a net DMI.
However, the local inversion symmetry breaking in a single FGT layer combined with a low interlayer coupling could give rise to a measurable DMI term \cite{lucassen2020stabilizing,walsem2020layer}, which could be interesting to investigate more specifically in further research. 

At this point we would like to note, that besides the counterclockwise rotating N\'eel spin spiral an additional spin texture is simultaneously present in the experiments, where the magnetization rotates mainly in the $xy$-plane.
The SEMPA measurements are depicted in the supplementary Information Section SIII and both a clockwise and counterclockwise rotation of this spin texture is observed.
We suspect that local fluctuations in strain or Fe atom concentration deficiency caused variations in the magnetic parameters (e.g. magnetic anisotropy and DMI), allowing both the out-of-plane and in-plane spin textures to stabilize \cite{doi:10.1021/acs.nanolett.9b03316,PhysRevB.93.014411}.
A qualitative agreement between micromagnetic simulations including a positive DMI and these SEMPA measurements is found and discussed in Supplementary Section SIV. 

\begin{figure}
\centering
\includegraphics{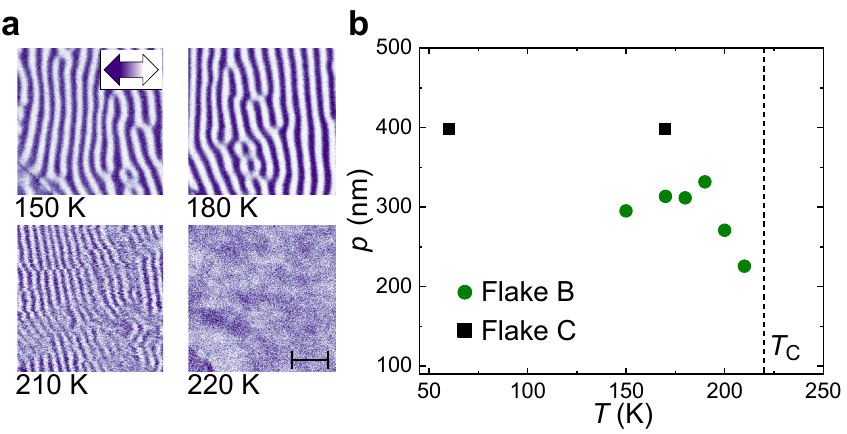}
\caption{\label{fig:figure3} Temperature dependence of the magnetic texture on the surface of FGT. \textbf{a)} SEMPA images showing the $m_\mathrm{x}$ contrast for $T=150$, $180$, $210$ and $220$~\si{K} on flake B. The in-plane magnetization direction is indicated by the arrow in the top left-hand image. The scale bar in the bottom right-hand image indicates $1$~\si{\mu m} and holds for all images. \textbf{b)} The period of the magnetization in $m_\mathrm{x}$ is plotted as a function of temperature for flake B and C and $T_C$ is indicated by the dashed line. 
}
\end{figure} 

Finally, we turn our attention to the temperature dependence of the magnetic texture.
\figref{fig:figure3}a shows $m_\mathrm{x}$ SEMPA images for the same area on a different flake (flake B) at various temperatures.
A similar magnetization pattern to the one discussed previously is observed for temperatures below the Curie-temperature of FGT (e.g. $T=150$~\si{K} and $180$~\si{K}).
The periodicity of the spin texture is almost independent of temperature in this range, and shows a period ($p$) of approximately $300$~\si{nm} for Flake B, which is plotted in \figref{fig:figure3}b.
Upon increasing the temperature towards $T_C$ the period of the magnetization rapidly decreases, showing a period of approximately $225$~\si{nm} at $T=210$~\si{K}, where fluctuations in the magnetization pattern are also observed due to the local heating induced by the electron beam.
Above $220$~\si{K} the Curie-temperature is reached (indicated by the dashed line) and the magnetic contrast completely vanishes.
A different FGT flake (flake C) was also investigated using a larger temperature range by cooling down the setup with liquid helium. 
We find that for flake C the period of the magnetic texture (here $p \approx$ 400~\si{nm}) also remains constant when varying the temperature from $60$~\si{K} to $170$~\si{K}.

The stability of the magnetic texture is unexpected in such a wide temperature range because of the strong temperature dependence of the anisotropy discussed earlier.
Although this temperature dependence of the anisotropy is shown in thin FGT flakes of 10~\si{nm} \cite{Tan2018HardFe3GeTe2}, we assume it to be present in the several $100$~\si{nm} thick flakes studied here as well.
A similar temperature dependence of the anisotropy has been measured in other bulk van der Waals materials, like Cr$_2$Ge$_2$Te$_6$ \cite{Zhang_2016}. 
The constant period in the magnetic texture therefore indicates that the change of the anisotropy upon increasing the temperature does not seem to influence the magnetic texture.
This implies either that other magnetic parameters, such as the magnetic exchange or DMI, change in a similar way with temperature as the anisotropy, resulting in no net change on the spin spiral period.
On the other hand, the constant period might indicate that the anisotropy contribution is small compared to the other magnetic terms. 
In the latter case we estimate the DMI to be at least $D>0.2$~\si{mJ ~m^{-2}} as explained above. 

In summary, we have investigated the magnetic texture in the top layer of FGT using SEMPA.
Our measurements revealed the presence of out-of-plane spin spirals rotating in a counterclockwise fashion, which indicates the presence of a positive DMI in FGT, although the origin of the DMI remains elusive, with a possible explanation being a local inversion symmetry breaking in single FGT layers.
We find the spin spiral pattern to be nearly temperature independent, indicating that the magnetic structure is not dominated by the anisotropy, or that other magnetic parameters have similar temperature dependencies.
Our work provides an important starting point for the use of (bulk) magnetic van der Waals materials for chiral magnetism.
We note that the value for the DMI estimated here for bulk FGT could possibly be further increased by enhancing the spin-orbit interaction through proximity effects from other vdW materials \cite{ZUTIC201985}, similarly to what is done in sputtered thin metallic layers such as Pt/Co systems.
The demonstration of chiral magnetic structures at the surface of bulk vdW materials is a crucial step towards more complicated vdW heterostructures with engineered magnetic properties.
 
\section{Acknowledgements}
This work is part of the research programme Exciting Exchange with project number 14EEX06, which is (partly) financed by the Dutch Research Council (NWO). M.H.D.G. acknowledges financial support from NWO (Veni 15093) and this project has received funding from the European Research Council (ERC) under the European Union's Horizon 2020 research and innovation programme (grant agreement No. 725509). 

\section{Author Contributions}
M.J.M. and M.H.D.G. conceived the idea and initiated the project. M.J.M. prepared the samples, performed the experiments and data analysis with M.H.D.G.'s assistance. M.J.M. performed the micromagnetic simulations with assistance from J.L.. R.A.D., H.J.M.S., B.K., R.L. and M.H.D.G. supervised the project. M.J.M. wrote the manuscript with assistance from M.H.D.G.. All authors commented on the final version of the manuscript.

\section{Methods}
\subsection{Sample preparation}
Our samples were mechanically exfoliated of a bulk FGT crystals (HQ Graphene) onto a Si wafer. This was done in high vacuum, at a pressure lower than 10$^{-7}$~\si{mbar} to avoid oxidation of the exfoliated FGT crystals. A dusting layer of $0.3$~\si{nm} cobalt was deposited by DC sputter deposition. The base pressure of the system was $4\times 10^{-9}$~\si{mbar} and the Ar pressure during deposition was $1\times 10^{-2}$~\si{mbar}. After the deposition the sample was transported to the SEMPA setup \textit{in-situ} and kept in UHV conditions with a base pressure of $2\times 10^{-10}$~\si{mbar}. The sample stage of the SEMPA setup was cooled down with liquid nitrogen or helium, resulting in a lowest reachable temperature of $140$~\si{K} and $60$~\si{K}, respectively.

We found no measurable effect of the Co dusting layer on the spin spiral periodicity. In the Supplementary Information Section SII we provide a comparison of SEMPA images of the same region of a FGT flake without and with Co dusting at low and room temperatures.

\subsection{Measurements}
The FGT flakes are always zero-field cooled, since we are not able to apply any magnetic fields in our SEMPA setup. A heater close to the sample stage allows us to measure at intermediate temperatures. In the SEMPA setup we are able to map the in-plane magnetization vector and additionally gain information on the out-of-plane magnetization by tilting the sample \cite{Lucassen2017ScanningTextures}. This results in the projection of the magnetization on the in-plane measurement axis, which is adjustable and well-defined for the $m_\mathrm{y}$ image as depicted schematically in \figref{fig:figure1} c and d. In the $m_\mathrm{x}$ image the out-of-plane projection depends strongly on the sample mounting and flake attachment to the substrate. 

\bibliography{references}
\end{document}


\title{Supplementary Information: Chiral Spin Spirals at the Surface of van der Waals Ferromagnet Fe$_3$GeTe$_2$}
\author{Mariëlle J. Meijer}
\email{m.j.meijer@tue.nl}
\affiliation{Department of Applied Physics, Eindhoven University of Technology, P.O. Box 513, 5600 MB Eindhoven, the Netherlands}
\author{Juriaan Lucassen}
\affiliation{Department of Applied Physics, Eindhoven University of Technology, P.O. Box 513, 5600 MB Eindhoven, the Netherlands}
\author{Rembert A. Duine}
\affiliation{Department of Applied Physics, Eindhoven University of Technology, P.O. Box 513, 5600 MB Eindhoven, the Netherlands}
\affiliation{Institute for Theoretical Physics, Utrecht University, Leuvenlaan 4, 3584 CE Utrecht, the Netherlands}
\author{Henk J.M. Swagten}
\affiliation{Department of Applied Physics, Eindhoven University of Technology, P.O. Box 513, 5600 MB Eindhoven, the Netherlands}
\author{Bert Koopmans}
\affiliation{Department of Applied Physics, Eindhoven University of Technology, P.O. Box 513, 5600 MB Eindhoven, the Netherlands}
\author{Reinoud Lavrijsen}
\affiliation{Department of Applied Physics, Eindhoven University of Technology, P.O. Box 513, 5600 MB Eindhoven, the Netherlands}
\author{Marcos H. D. Guimar\~aes}
\email{m.h.guimaraes@rug.nl}
\affiliation{Department of Applied Physics, Eindhoven University of Technology, P.O. Box 513, 5600 MB Eindhoven, the Netherlands}
\affiliation{Zernike Institute for Advanced Materials, University of Groningen, Nijenborgh 4, 9747 AG Groningen, the Netherlands}

\date{\today}
\maketitle

\section{SEM and AFM measurements on FGT}
\begin{figure}
\centering
\includegraphics[width=\textwidth]{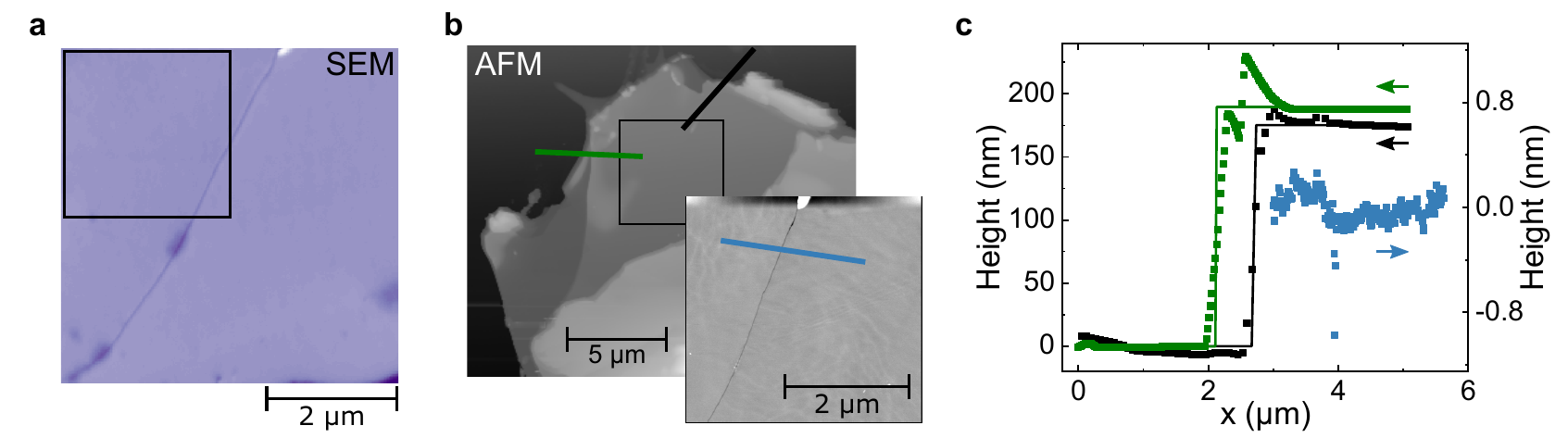}
\caption{\label{fig:AFM} \textbf{a} SEM image measured with SEMPA. 
The area indicated by the black outline was discussed in the main paper in Fig. 1 and the magnetic contrast in the full area is shown in \figref{fig:figure3}. 
\textbf{b} AFM images of the sample discussed in the main paper (Fig. 1). 
An overview scan is depicted as well as a detailed scan (inset) of the area measured with SEMPA. 
\textbf{c} Height profiles of the flake with the location of the line traces indicated in \textbf{b}.  }
\end{figure}

In this section we show additional SEM and AFM images of flake A. 
In \figref{fig:AFM}a a SEM image  measured with the SEMPA system  is depicted. 
In SEMPA four channeltrons collect electrons scattered from a W(001) crystal at specific diffraction spots and a magnetic contrast is obtained by subtracting the counts from channeltrons situated on opposite sides \cite{doi:10.1063/1.3534832}. 
When all the counts of the different channeltrons are added up we discard the spin information and are left with a SEM image as depicted here. The resolution obtained in this SEM image is identical to the one of the SEMPA images in the main paper (Fig. 1) and features of $30$~\si{nm} can easily be distinguished. 
The area studied in the main paper (Fig. 1) is outlined in black and the SEMPA images corresponding to the full area are discussed in 
\figref{fig:figure3}. 

In \figref{fig:AFM}b an AFM scan of a large area of the flake is shown. 
The area measured with SEMPA is outlined in black and the inset shows a detailed scan. The black, green and blue line indicate at which positions the height profiles in \figref{fig:AFM}c are measured. 
The green and black height profile use the $y$-axis on the left hand side and measure a flake thickness of $185 \pm 7$~\si{nm}.
The blue height profile is taken across the area measured with the SEMPA and the $y$-axis on the right hand side is used. 
Here, the scale of $0.8$~\si{nm} corresponds to the thickness of a single FGT layer. 
Within this height profile and specifically across the narrow crack visible in \figref{fig:AFM}a and b, we find no height step in the flake. 

\section{Magnetic contrast in SEMPA images} 
\begin{figure}
\centering
\includegraphics[width=0.8\textwidth]{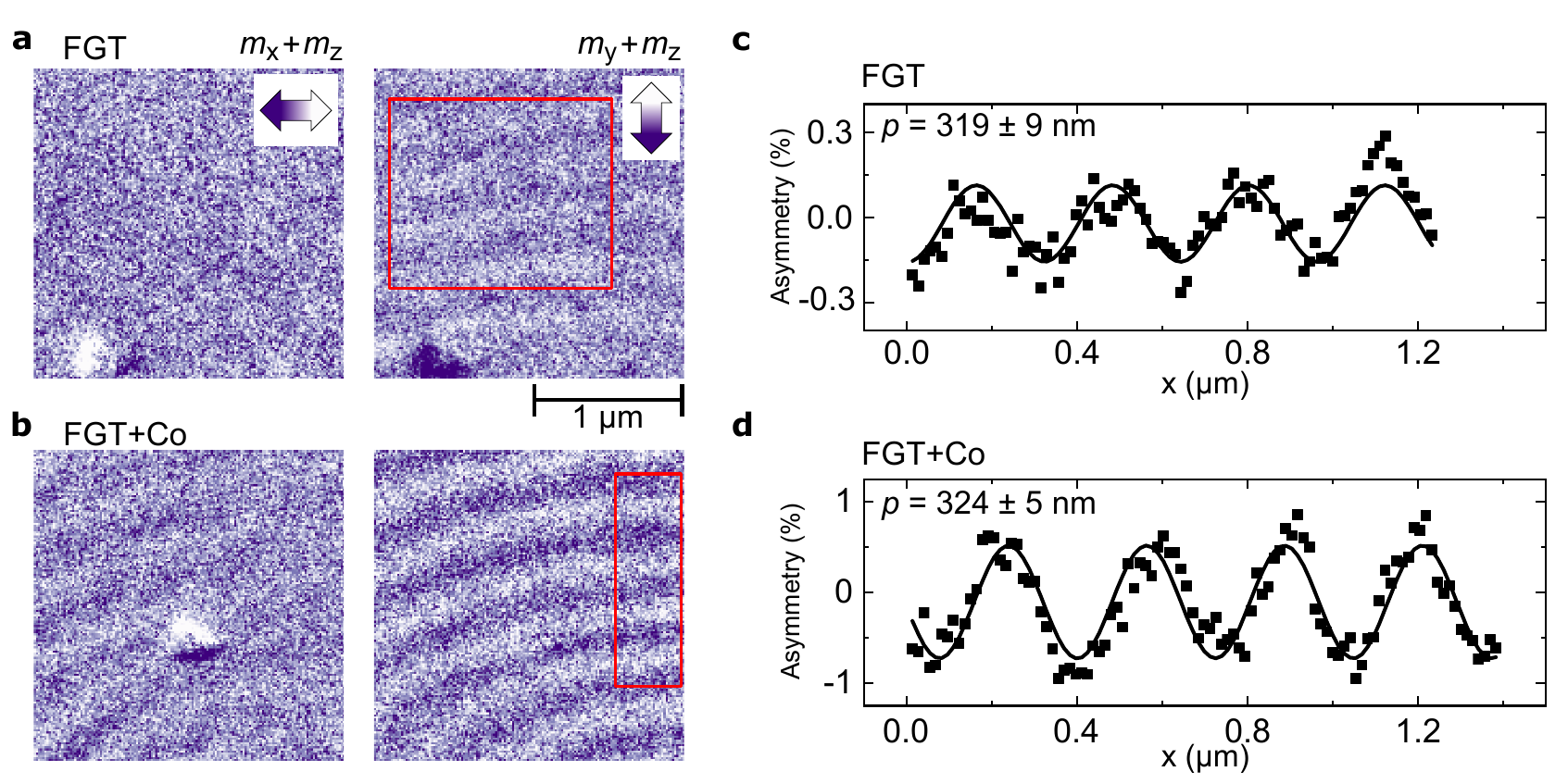}
\caption{\label{fig:figure1} \textbf{a} SEMPA images of FGT with the horizontal and vertical in-plane magnetization depicted in the left and right image, respectively. 
\textbf{b} SEMPA images of the same area after 0.3 nm of Co is deposited on FGT. \textbf{c,d} Average magnetic contrast data along the vertical direction of the red rectangle indicated in the SEMPA images. 
The data are fitted with a sinusoidal function (black line) and a similar periodicity $p$ is extracted. The defects in panel a and b are the same. }
\end{figure}

In this section we discuss the differences in magnetic contrast that can be found in the SEMPA images. 
First, we depict in \figref{fig:figure1} that the magnetic contrast in the SEMPA images is greatly enhanced by depositing a thin layer of 0.3 nm of Co on FGT. 
The SEMPA images of pure FGT are depicted in \figref{fig:figure1}a and only a very faint contrast in the right SEMPA image, containing the $m_\mathrm{y}$-contrast, is visible. 
In \figref{fig:figure1}b the same area is imaged after 0.3 nm of Co is deposited on FGT (see main paper for details). 
Here, the defect in the lower left corner of \figref{fig:figure1}a corresponds to the defect in the middle of the image of \figref{fig:figure1}b. 
The magnetic contrast in \figref{fig:figure1}b is greatly enhanced and a clear magnetic signal is found showing horizontal lines.
In the right SEMPA images of \figref{fig:figure1}a and b the magnetic contrast is further analyzed in \figref{fig:figure1}c and d, where the data in the red highlighted area is averaged along the vertical direction.
In both graphs a periodic behavior is found, but the amplitude of the oscillations is much stronger in \figref{fig:figure1}d due to the added Co. 
We fit both data sets with a sinusoidal function (black lines) and find a similar periodicity of the magnetic signal, which indicates that the thin layer of Co does not induce a magnetic pattern on its own but rather follows the magnetic texture of FGT. 
This statement is confirmed by the fact that as soon as the Curie-temperature of FGT is reached, the magnetic contrast vanishes entirely (see Fig. 3a of the main paper). 

\begin{figure}
\centering
\includegraphics[width=0.6\textwidth]{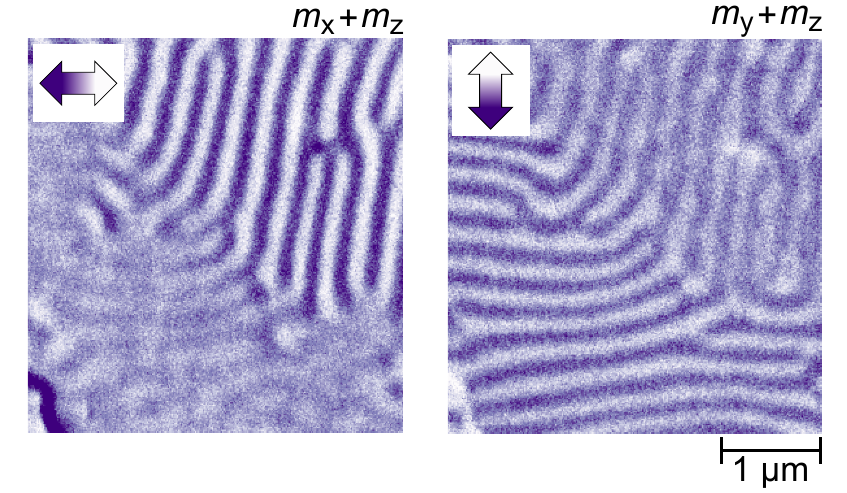}
\caption{\label{fig:figure2} SEMPA images of FGT with the horizontal and vertical in-plane magnetization depicted in the left and right image, respectively. 
The area in the upper right hand corner shows a vertically aligned magnetic texture and a strong magnetic contrast is found in the left SEMPA image. 
In the bottom left hand corner of the image the magnetic texture is aligned horizontally and only in the right SEMPA image a strong magnetic contrast is measured.  }
\end{figure}

In the second part we discuss the differences in magnetic contrast within SEMPA images. 
A clear example is depicted in \figref{fig:figure2}, where a pristine area of an FGT flake is depicted. 
The left SEMPA image ($m_\mathrm{x}$ contrast) shows a very strong magnetic contrast in the upper right hand corner of the image and almost no magnetic contrast in the bottom left hand corner. 
In the right SEMPA image ($m_\mathrm{y}$ contrast) the opposite is observed. 
From this figure we find that in FGT a strong in-plane magnetization contrast in $m_\mathrm{x}$ or $m_\mathrm{y}$ corresponds to a magnetic pattern aligned either vertically or horizontally. 

The $m_\mathrm{z}$ component, which might be present in both SEMPA images, can only be adjusted in the $m_\mathrm{y}$ SEMPA image when rotating the sample stage. 
In order to obtain as much information of the SEMPA images as possible we study in particular the vertically aligned magnetic patterns. Here, the main component of the in-plane magnetization will be present in the $m_\mathrm{x}$ SEMPA image. 
A small $m_\mathrm{z}$ component in this image can not be differentiated due to the higher sensitivity of the SEMPA for in-plane components. 
The other SEMPA image then consists of a  (small) $m_\mathrm{y}$ and $m_\mathrm{z}$ magnetization contribution, which can be separated by rotating the sample (see main paper for details). 

\section{In-plane spin spiral rotating CW and CCW}
\label{in-plane}

\begin{figure}
\centering
\includegraphics{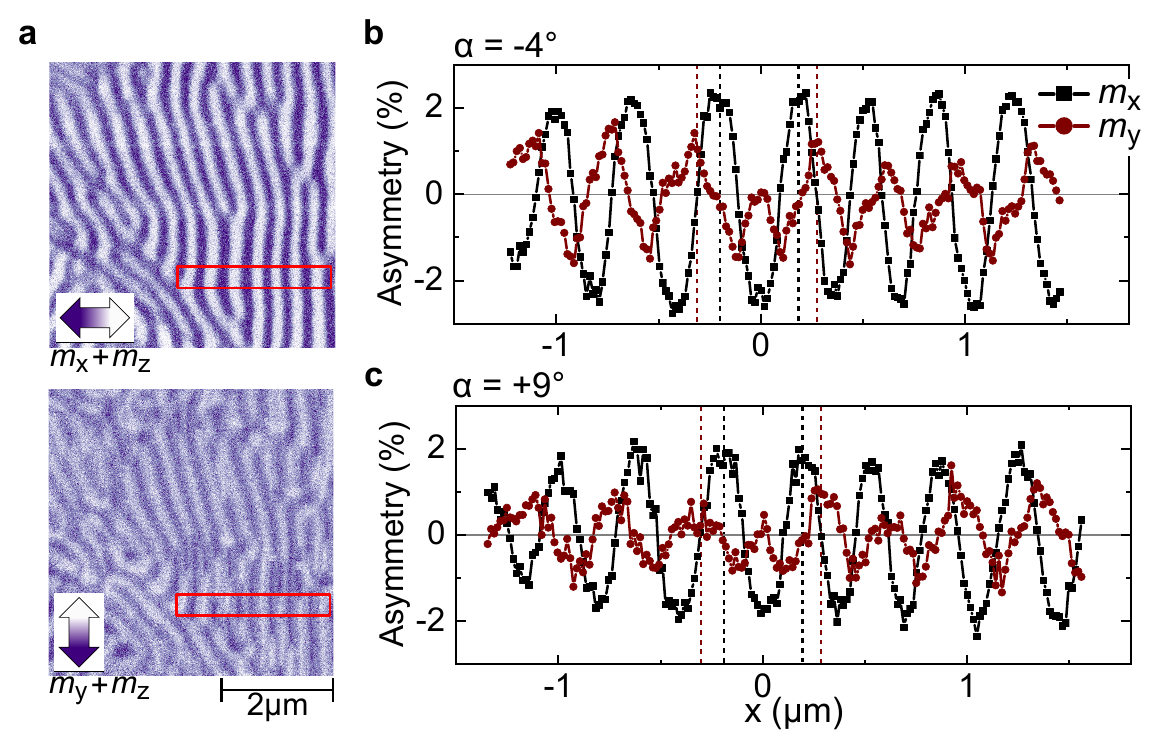}
\caption{\label{fig:angle} \textbf{a} SEMPA images of flake B at $\alpha=-4^\circ$ with the horizontal and vertical in-plane magnetization depicted in the top and bottom image, respectively. 
The red area is averaged along the vertical axis in \textbf{b}. 
A phase shift of $-\pi/2$ and $+\pi/2$ is observed between the $m_\mathrm{x}$ and $m_\mathrm{y}$ SEMPA data for negative and positive $x$. In \textbf{c} $\alpha=+9^\circ$ and the phase shift remains the same as in \textbf{b}.  
}
\end{figure}

In this section we consider a different magnetic spin texture to the one presented in the main paper. Here, an in-plane rotating spin texture on the surface of FGT is discussed rotating in a clockwise as well as a counterclockwise fashion.
We first focus on measurements on flake B depicted in \figref{fig:angle}a.
The in-plane magnetization in the $m_\mathrm{x}$ direction is shown in the top SEMPA image and for $m_\mathrm{y}$ in the bottom SEMPA image. 
The area that is investigated in more detail is highlighted by the red outline and the averaged data in this area is plotted in \figref{fig:angle}b.
A constant sinusoidal magnetic texture is obtained for the black data, corresponding to the $m_\mathrm{x}$ SEMPA image. 
The red data ($m_\mathrm{y}$ SEMPA image), however, shows an triangular shaped oscillatory behavior, which is shifted by a phase of $-\pi/2$ for negative $x$ values and $+\pi/2$ for positive $x$ values with respect to the $m_\mathrm{x}$ signal.
We find that the sign of these phase shifts remains the same when $\alpha$ is rotated from $\alpha=-4^\circ$ (\figref{fig:angle}b) to $\alpha=+9^\circ$ (\figref{fig:angle}c). 
As discussed in the main paper, this indicates that the red data points in \figref{fig:angle}b correspond mainly to the in-plane magnetization signal ($m_\mathrm{y}$), rather than the out-of-plane component.
We therefore observe an in-plane rotating spin texture with no apparent preferred sense of rotation at the surface of FGT.
Moreover, the triangular shape of the $m_\mathrm{y}$ signal indicates, that the full magnetic texture is probably of a three-dimensional nature since the combination of the in-plane magnetic components do not result in a uniform magnetization.

\begin{figure}
\centering
\includegraphics{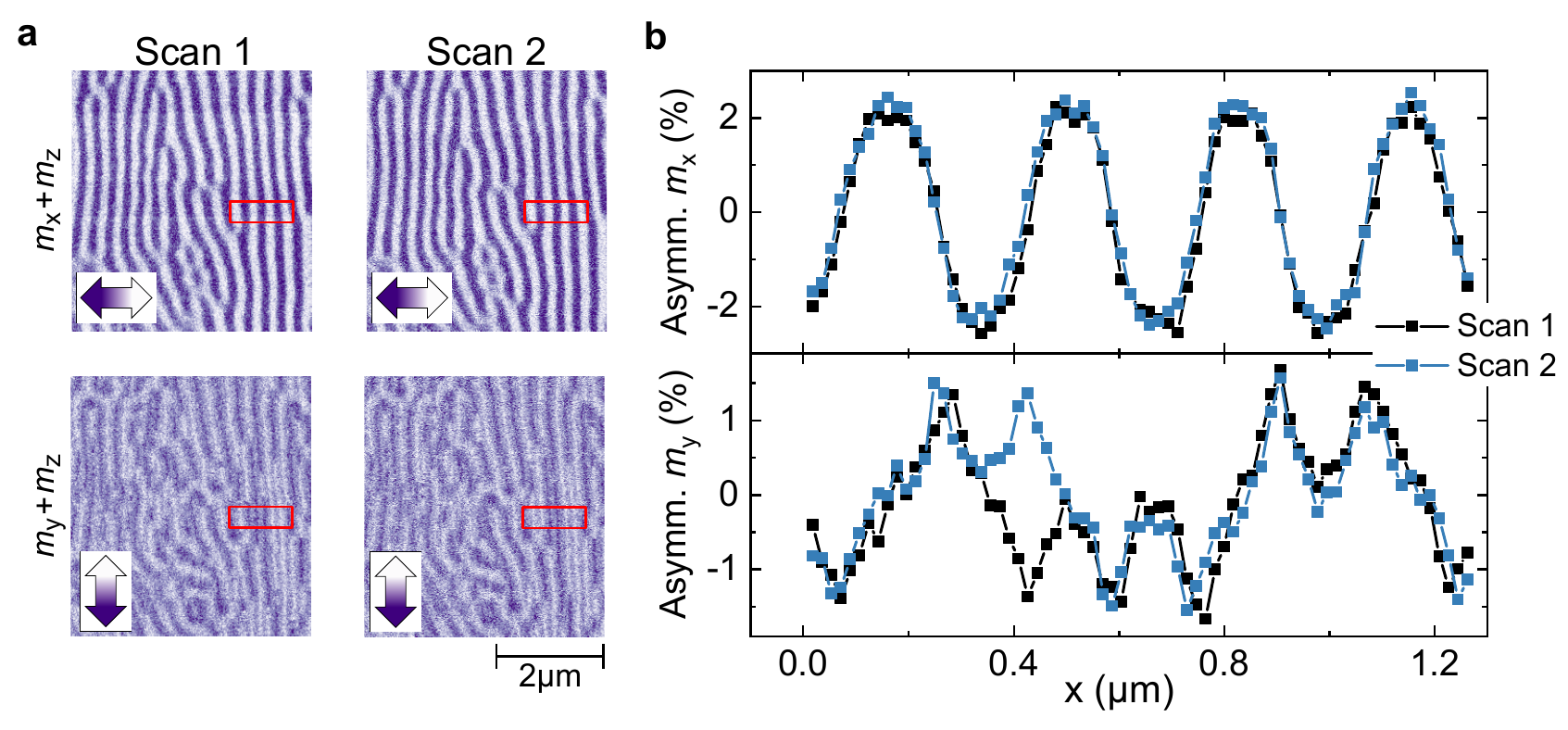}
\caption{\label{fig:change} \textbf{a} SEMPA images of two different SEMPA scans with the in-plane magnetization in the vertical direction and horizontal direction in the upper and lower row, respectively. \textbf{b} Averaged data for the $m_\mathrm{x}$ and $m_\mathrm{y}$ scan in the upper and lower panel, respectively. 
The signal in $m_\mathrm{x}$ is identical in both scans, but for the $m_\mathrm{y}$ scan the the magnetization changed sign around $x=0.4$~\si{\mu m}. }
\end{figure}

The rotation direction of the spin texture can vary quickly in space and time, as can be seen in \figref{fig:change} in flake B. 
In \figref{fig:change}a we show two SEMPA scans measured a few minutes apart, where the horizontal and vertical in-plane magnetization are depicted in the upper and lower images, respectively.
The data in the red area is averaged and plotted in \figref{fig:change}b. 
The $m_\mathrm{x}$ signal shows a constant sinusoidal behavior, whereas the $m_\mathrm{y}$ shows a clear triangular shaped oscillatory behavior.  
At irregular intervals the phase shift between the $m_\mathrm{x}$ and $m_\mathrm{y}$ signal changes from $+\pi/2$ to $-\pi/2$. Moreover, in between two scans the phase shift changes locally, as can be seen around $x=0.4$~\si{\mu m}. 
The simultaneous presence of both the clockwise and counterclockwise rotating in-plane spin texture and that the magnetic texture can change from one into the other over time indicate, that they are energetically similar at the surface of FGT.

\begin{figure}
\centering
\includegraphics{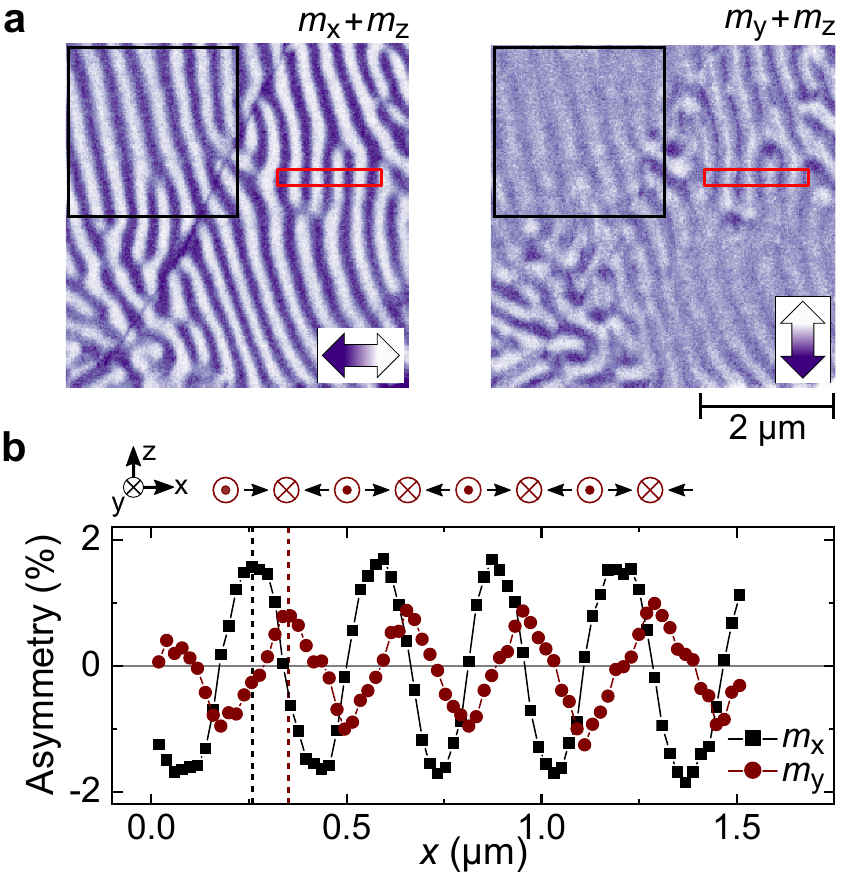}
\caption{\label{fig:figure3} \textbf{a} SEMPA image of flake A with the horizontal and vertical in-plane magnetization depicted in the left and right image, respectively. 
The area outlined in black is discussed in the main paper (Fig. 1). 
A different magnetization pattern becomes apparent in the upper right hand corner (mainly visible in $m_\mathrm{y}$). 
The data outlined in red is plotted in \textbf{b} for the $m_\mathrm{x}$ (black data) and $m_\mathrm{y}$ (red data) magnetization. 
A schematic of the rotating magnetization in the $\mathrm{xy}$-plane is illustrated on top of the panel.
}
\end{figure}

Lastly we have a look at SEMPA measurements of flake A, depicted in \figref{fig:figure3}a. 
In the main paper in Fig. 1 the area indicated by the black outline is discussed.
Here, we focus on the area of the image in the upper right hand corner. In this region the same in-plane rotating magnetic texture is found as discussed in this section and the data outlined in red is plotted in detail in \figref{fig:figure3}b. 
From this measurement we find that the in-plane rotating spin textures can be present simultaneously with the counterclockwise Néel spin spirals (main paper Fig. 1) at the surface of bulk FGT.

\section{Qualitative micromagnetic simulations}

The micromagnetic simulations in Fig. 2 of the main paper and this section are obtained with \mumax \cite{Vansteenkiste2014}. 
For all simulations the following settings were used. 
The cell sizes were (0.4, 8, 0.8) nm for $(x,y,z)$ with periodic boundary conditions in the $x$- and $y$-direction for 32 repeats.
The cell size in $z$ corresponds to the height of a single FGT layer and in total 128 layers were simulated. 
A saturation magnetization of $M_\mathrm{S}=0.38$~\si{MA m^{-1}} and an exchange stiffness of $A=1$~\si{pJ m^{-1}} were used \cite{De_Clercq_2017}.
Additionally an interlayer exchange interaction was added with the strength of 10\% of the exchange stiffness. 
We implemented this interaction via the RKKY interaction method discussed elsewhere \cite{De_Clercq_2017}. 
The anisotropy was varied from $K=0.01-1.5$~\si{MJ m^{-3}} and for every layer an interfacial DMI was added with a strength between $D=0-0.8$~\si{mJ m^{-2}}. 

A magnetic domain texture was initialized in the following way: in total four alternating 'up' and 'down' domains were formed with 
$(m_\mathrm{x},m_\mathrm{y}, m_\mathrm{z})=(0.408, 0.408,\pm 0.816)$ and in between these domains 5~\si{nm} domain walls are initialized with $(m_\mathrm{x},m_\mathrm{y}, m_\mathrm{z})=(0.667,0.667,0.333)$. 
The simulations are minimized (using default settings) from the initialized state to obtain the equilibrium magnetization. 
For different simulation sets the signs of the initialized in-plane components of domains and/or domain walls are changed to check the consistency of the simulations. 
In Fig. 2c of the main paper the simulations shows a Bloch wall pointing in the $+y$-direction (blue color). 
However, a Bloch wall in the $-y$-direction (yellow color) would have the same energy. 
Both Bloch wall configurations are found in different simulations sets. 
This is not the case for Fig. 2d, however, where a finite D is present. For the indicated parameters all simulation sets show the same result, namely a counterclockwise rotating spin spiral.

\begin{figure}
\centering
\includegraphics{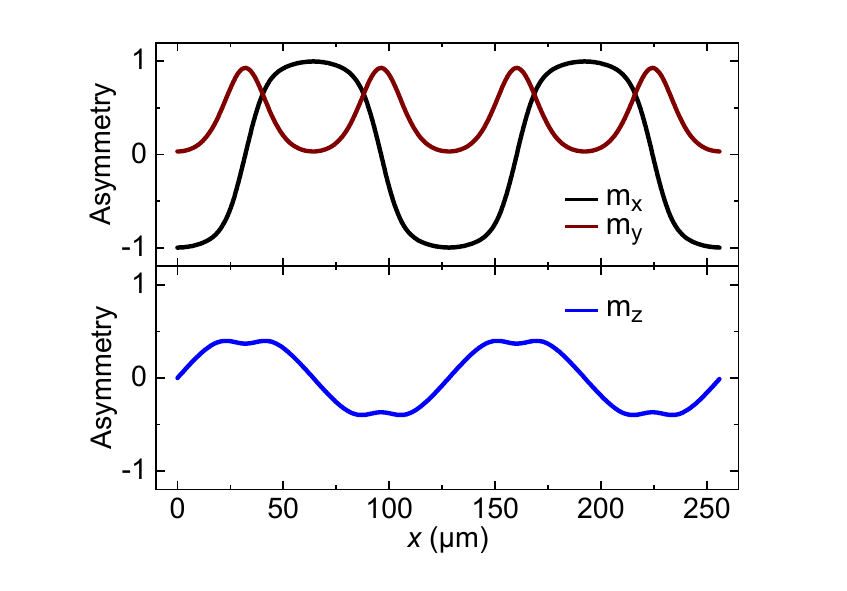}
\caption{\label{fig:mumax} Micromagnetic simulation of the surface magnetization of FGT for $D=0.2$~\si{mJ m^{-2}} and $K=40$~\si{kJ m^{-3}}. In the top panel the in-plane magnetic components are shown and the out-of-plane component in the lower panel. 
}
\end{figure}

Lastly we discuss the simulation results depicted in \figref{fig:mumax} that are in qualitative agreement with the in-plane magnetization texture discussed in Supplementary Section \ref{in-plane}. 
Here, the DMI is lower compared to the previous simulations, namely $D=0.2$~\si{mJ m^{-2}} and $K=40$~\si{kJ m^{-3}} and the upper panel of the figure shows the simulated magnetization profile of the top layer of FGT in the $m_\mathrm{x}$ and $m_\mathrm{y}$ direction in black and red, respectively. 
A sinusoidal pattern is found for $m_\mathrm{x}$ and the $m_\mathrm{y}$ signal peaks at every zero transition of $m_\mathrm{x}$. 
The same behavior but for negative $m_\mathrm{y}$ values is found for a different simulation set. 
This magnetization profile (with either positive or negative $m_\mathrm{y}$ values) closely resembles the SEMPA images depicted in \figref{fig:change}b.  
As was indicated, the overall structure is expected to be three dimensional and the simulated $m_\mathrm{z}$ component is plotted in the lower panel of \figref{fig:mumax}. 

Overall we find from Fig. 2c and d of the main paper and \figref{fig:mumax} that we are able to qualitatively simulate the different magnetic textures observed at the surface of FGT. 
However, the low anisotropy values and origin of a DMI remain elusive. 
A better understanding of several of the magnetic parameters and their thickness and temperature dependence would allow us to match the measured periodicity of the spin spiral to simulations and moreover to predict the magnetic texture of FGT. 

\bibliography{references}